\documentclass[prd,twocolumn,showpacs,preprintnumbers,nofootinbib,amsmath,amssymb,nobalancelastpage]{revtex4}
\usepackage[pdftex]{graphicx}
\usepackage{latexsym,amsmath,amssymb,lmodern,float,url}
\usepackage{natbib}
\usepackage[pdftex,bookmarks,linktocpage,pdfpagelabels,plainpages=false,hyperfigures,linkcolor=blue,citecolor=blue,urlcolor=blue]{hyperref} \usepackage{xspace}
\usepackage{courier}
\hypersetup{colorlinks=true}
\usepackage{bm}
\usepackage{dcolumn}
\usepackage{amsmath}
\usepackage{amssymb}
\usepackage{color}
\usepackage{xcolor}
\usepackage{soul}
\usepackage{url}
\usepackage{float}

\usepackage{aas_macros}
\usepackage[normalem]{ulem}

\def\Eq#1{Eq.~(\ref{#1})}
\def\Fig#1{Fig.~(\ref{#1})}
\def\Sec#1{Sec.~\ref{#1}}

\def\H0{$H_0$}
\def\si8{$\sigma_8$}

\graphicspath{{./figs/}}

\begin{document}

\title{Cosmological constraints on late-universe decaying dark matter as a solution to the $H_0$  tension }

\author{Steven~J.~Clark} 
\email{steven$\_$j$\_$clark@brown.edu}
\affiliation{ Department of Physics, Brown University, Providence, RI 02912-1843, USA}
\affiliation{ Brown Theoretical Physics Center, Brown University, Providence, RI 02912-1843, USA}

\author{Kyriakos Vattis}
\email{kyriakos$\_$vattis@brown.edu}
\affiliation{ Department of Physics, Brown University, Providence, RI 02912-1843, USA}
\affiliation{ Brown Theoretical Physics Center, Brown University, Providence, RI 02912-1843, USA}

\author{Savvas M. Koushiappas}
\email{koushiappas@brown.edu}
\affiliation{ Department of Physics, Brown University, Providence, RI 02912-1843, USA}
\affiliation{ Brown Theoretical Physics Center, Brown University, Providence, RI 02912-1843, USA}

\date{\today}

\begin{abstract}

It has been suggested that late-universe dark matter decays can alleviate the tension between  measurements of $H_0$ in the local universe and its value inferred from cosmic microwave background fluctuations. It has been suggested that decaying dark matter can potentially account for this discrepancy as it reshuffles the energy density between matter and radiation and as a result allows dark energy to become dominant at earlier times.  In this work, we show that the low multipole amplitude of the cosmic microwave background anisotropy power spectrum severely constrains the feasibility of late-time decays as a solution to the $H_0$ tension. 
\end{abstract}

\maketitle

\section{Introduction}
The standard $\Lambda$CDM model has been established during the past decades as the standard cosmological model consisting of $70\%$ dark energy in the form of a cosmological constant $\Lambda$, $25\%$ cold dark matter (CDM) and $5\%$ baryonic matter. It has been very successful at describing the evolution of the Universe by accounting for a large range of observations, from cosmological scales (Cosmic Microwave Background (CMB) measurements \cite{2018arXiv180706209P}, Baryon Acoustic Oscillations (BAO)\cite{2019MNRAS.483.4866A}, redshift space distortions \cite{2018MNRAS.477.1604G}) to  galactic rotation curves \cite{2000MNRAS.311..441C} and galaxy cluster dynamics \cite{2008ApJ...679.1173R}. Despite the success of $\Lambda$CDM, as experimental measurements have improved, two prominent tensions have arisen. The first is the Hubble tension between early time cosmology with Cosmic Microwave Background \cite{2018arXiv180706209P} measurements and local late time cosmology from Type Ia Supernova \cite{2018ApJ...861..126R, 2019arXiv190307603R}. The second is the early \cite{2018arXiv180706209P} and late \cite{2020arXiv200211124D} cosmic variance measurements of the matter density field characterized by the value of \si8. 

The  discrepancy between the CMB measurement of \H0 and the distance ladder estimates from SNIa calibrated primarily using Cepheid stars evolved in the last few years from $2.5\sigma$ \cite{2016ApJ...826...56R} to $4.4\sigma$ \cite{2019arXiv190307603R}. While the ladder is a direct measurement of the expansion rate of the Universe today, CMB estimations are model dependent, having to extrapolate present-day values from a cosmological model that fits the CMB power spectra at the redshift of recombination. That is the reason why this tension is so important: it could potentially be an indication of new physics and thus  deviations from the standard $\Lambda$CDM cosmological model. 

As with any tension, multiple probes are needed to help clarify the origin of the observed discrepancy. Such an additional probe is the improved inverse distance ladder measurement by the Dark Energy Survey (DES)  \cite{2018arXiv181102376M}. In this case, the distances of SNIa are calibrated using BAOs, and the deduced value of $H_0$ is found to be consistent with the  measurements inferred directly from the CMB  \cite{2018arXiv180706209P}. The recent results from the Atacama Cosmology Telescope \cite{2020arXiv200707288A} confirmed the Planck measurements leaving little room for instrumental systematic errors.  In contrast, an independent inverse distance ladder measurement using quasars as an anchor  by H0LiCOW \cite{2019arXiv190704869W} is in agreement with the local measurement \cite{2019arXiv190307603R}, fuelling the tension between early and late time universe.  Yet another independent measurement  of \H0  was made possible based on the tip of the red giant branch \cite{2019ApJ...882...34F} finding an \H0 value laying midway in the range defined by the current Hubble tension. Similar mid-range value was obtained using gravitational waves produced from a binary neutron star merger \cite{2017Natur.551...85A,2017PhRvL.119p1101A}. Such gravitational wave ``standard siren" measurements of $H_0$ are extremely important because they do not rely on light, and they are governed by different systematic errors, though the observation of more events is needed to reduce the uncertainty to the percent level \cite{2018arXiv180705667F,2018arXiv181111723M,2018arXiv180203404F,2018arXiv180610596H,2018Natur.562..545C,2018PhRvL.121b1303V}. 

The origin of this discrepancy is still under debate.  Potential systematics at play were claimed as an explanation \cite{2018arXiv181002595S,2018arXiv181003526R,2018arXiv181202333V,2018arXiv181004966B}, however recently it was shown that the tension exists between all late and early universe datasets at high significance ~\cite{2019NatRP...2...10R} regardless of the dataset used.
There have been multiple attempts to relieve the data tension by introducing new physics and extensions to  $\Lambda$CDM by modifying either the behavior of dark energy or dark matter. The work of Knox and Millea \cite{2020PhRvD.101d3533K} points towards early universe solutions to be the less unlikely but such solutions fail to be in agreement with large scale structure observations as shown in \cite{2019arXiv190707953V,2020arXiv200611235I,2020arXiv201004158J}.

Dark energy modifications to the standard cosmological model include  a negative cosmological constant model~\cite{2018arXiv180806623D} though later proven insufficient to solve the tension \cite{2019arXiv190707953V}, and a dynamical dark energy equation of state ~\cite{2018arXiv180902340G,2019arXiv190304865K}. Another promising proposal has been based on an early period of dark energy domination that changes the size of the acoustic horizon \cite{2018arXiv181104083P,2019arXiv191010739N,2019arXiv191111760S}, while others include vacuum phase transitions \cite{2018PhRvD..97d3528D,2018arXiv181011007B,Banihashemi:2018oxo}, interacting dark energy \cite{2019arXiv190804281D,2019MNRAS.482.1858Y, 2019MNRAS.482.1007Y, 2018JCAP...09..019Y}, as well as quintessence field models \cite{2018JCAP...09..025M,2020PhRvD.101l3516A}  and Axion Dark Energy \cite{2019arXiv191000459C}. 

Modifications to the dark matter sector include partially acoustic dark matter models \cite{2017PhRvD..96j3501R}, charged dark matter with chiral photons \cite{2017PhLB..773..513K}, dissipative dark matter models \cite{2019APh...105...37D}, cannibal dark matter \cite{2018PhRvD..98h3517B}, non-thermal dark matter \cite{2019arXiv191205563A}, and axions \cite{2018JCAP...11..014D}. Decaying dark matter models were also considered  especially because of their properties of solving some small scale structure formation problems  \cite{2019PhRvD..99l1302V, 2015PhRvD..92f1301A,2017JCAP...10..028B,2018PhRvD..98b3543B,Pandey:2019plg,1984MNRAS.211..277D,1985PAZh...11..563D,1988SvA....32..127D,Hooper:2011aj}. Finally, modifications to the general theory of relativity were also proposed \cite{2018arXiv180909390E,2017arXiv171009366K,2017JCAP...10..020R,2020arXiv200411161B,2020arXiv200414349B}. Nevertheless, none of the aforementioned models have been completely successful on relieving the tension.

The tension in the amplitude of the variance on scales of $8 h^{-1} {\mathrm{Mpc}}$,  $\sigma_8$, appears to be well defined in observations, \cite{2009ApJ...692.1060V,2013PhRvL.111p1301M,2015PhRvD..91j3508B,2015MNRAS.451.2877M,2016PhRvD..93d3522R,2018arXiv180706209P,2020arXiv200211124D,2018A&A...614A..13S,2019arXiv190105289D} however it is not as robust as the \H0 discrepancy since its significance varies only from $1.5\sigma$ to $2.5\sigma$ depending on which late time probe one compares with the CMB-derived estimates. Despite that, there have already been multiple attempts in the literature to address the tension. A quite popular topic has been the introduction of self interactions in the dark sector, most notably by introducing self interaction in dark energy  \cite{2016PhRvD..94d3518P, 2014PhRvL.113r1301S, 2014PhRvD..89h3517Y, 2017EL....12039001G,2018MNRAS.478..126G} in an attempt to erase structure in the late universe and relax the tension. Additionally dark radiation and dark matter self-interactions have been proposed \cite{2016PhLB..762..462K,2017PhLB..768...12K} trying to solve the problem in a similar manner while others take a different approach for example by introducing a model with dark matter-neutrino interactions \cite{2018PhRvD..97d3513D} or modifications to gravity \cite{2018PhRvD..97j3503K}. On the other hand, models invoking a viscous dark matter \cite{2017JCAP...11..005A}, an effective cosmological viscosity \cite{2018JCAP...05..031A} or neutrino self-interactions \cite{2019arXiv190200534K} attempt to solve both tensions simultaneously.

It has been proposed that decaying dark matter can be a possible solution to not only the Hubble tension~\cite{2019PhRvD..99l1302V} but also to the $\sigma_8$ controversy because it has the characteristic of erasing structure in the late universe, which is what is needed to save both problems. In general, constraints on decaying dark matter models have been constrained by various methods \cite{2013PhRvD..88l3515W, 1998MNRAS.296...44G, 2006ApJS..163...80M, 2013A&A...559A..85P, 2018JCAP...04..053A, 2014MNRAS.445..614W, 2015JCAP...09..067E, 2015PhRvD..92f1303B,2016PhRvD..94b3528C,2018PhRvD..97h3508C}.
In this work, we expand on the simplified treatment of the effects of decaying dark matter in \cite{2019PhRvD..99l1302V} to
the investigate the impact of a two-body decaying dark matter model on the power spectrum of the cosmic microwave background, specifically for decays that can alleviate the \H0 and \si8 tensions.  In  \Sec{sec:decaying_dm} we review the basic properties of two-body decays and its cosmological implications. In \Sec{sec:constraints} we describe  the CMB constraints of such a model, and we conclude in \Sec{sec:conclusion}.

\section{Decaying dark matter and cosmology} \label{sec:decaying_dm}
In this section we discuss the physical properties and cosmological characteristics of a two-body decaying dark matter scenario.
In the rest of the section, we assume the default parameters of our cosmology software, CLASS~\cite{2011JCAP...07..034B}, consistent with the best fit to the Planck 2013 + WP (WMAP Polarization) results~\cite{2014A&A...571A..16P}: the peak scale parameter $100\theta_{\rm s} = 1.042143$, the baryon density today $\Omega_{\rm b} h^2 = 0.022032$, the dark matter density today assuming a non decaying cosmology $\Omega_{\rm CDM} h^2 = 0.12038$, the redshift of reionization $z_{\rm reio} = 11.357$, the matter power spectrum value at pivot scale $A_{\rm s} = 2.215 \times 10^{-9}$, and the scalar tilt $n_{\rm s} = 0.9619$ where the pivot scale is $k=0.05$. These parameters were used both for demonstration of the properties of the decaying dark matter model as well as for the comparison with  $\Lambda$CDM.

\subsection{Two-body decays} \label{sec:decaying_dm_formulism} 
The decaying dark matter model we consider consists of a single cold unstable parent particle created in the early Universe which  decays into two daughter particles as $\psi \rightarrow \gamma^\prime + \chi$: one massless (e.g., a dark photon \cite{2003PhRvL..91a1302F, 2016PhRvD..94a5018C, 2018PhRvL.120v1102K}) and one massive particle. The model is characterized by only two parameters; the decay width $\Gamma$ and the fraction $\epsilon$ of rest mass energy of the parent particle transferred to the  massless particle $\gamma^\prime$. From here on, we use subscripts $0$, $1$, and $2$ corresponding to the parent, massless daughter, and massive daughter to identify quantities related to each species respectively. Following the work in Ref.~ \cite{2014PhRvD..90j3527B,2019PhRvD..99l1302V}, we can write the cosmological evolution of the densities of all species as
\begin{eqnarray}
\dot{\rho}_0&=& - 3\frac{\dot{a}}{a}\rho_0 -\Gamma \rho_0 \label{eq:rho0}\\
\dot{\rho}_1&=& - 4\frac{\dot{a}}{a}\rho_1 + \epsilon \Gamma \rho_0 \label{eq:rho1}\\
\dot{\rho_2} &=& - 3(1+w_2)\frac{\dot{a}}{a}\rho_2 + (1-\epsilon) \Gamma \rho_0 \label{eq:rho2}
\end{eqnarray}
where $\rho_i$ is the energy density of species $i$,  derivatives are with respect to time, and $a$ the scale factor. The quantity $w_2(a)$ is the dynamical equation of state of the massive daughter particle and it is given by (see~\cite{2014PhRvD..90j3527B}) 
\begin{eqnarray}
w_2(a)&=&\frac{1}{3} \frac{\Gamma \beta^2_2}{e^{-\Gamma t_\star}-e^{-\Gamma t}} \nonumber \\
&\times&  \int^a_{a_\star} \frac{e^{-\Gamma t_D} \, \, \mathrm{d}  \ln(a_D)}{H_D[(a/a_D)^2(1-\beta^2_2)+\beta^2_2]}. 
\label{equ:w_2}
\end{eqnarray}
where $\beta=\epsilon/(1-\epsilon)$ is the velocity in units of $c$ of  $\chi$ particles at production, and $t=t(a)$, the time that corresponds to scale factor $a$.  The constant $t_\star$ sets the initial conditions, $\rho_1(t=t_\star) = \rho_2(t=t_\star) = 0$ and $\rho_0(t=t_\star) = \rho_{\mathrm{crit}} \Omega_{\mathrm{DM}}$ with  $ \rho_{\mathrm{crit}}$ being the critical density and $\Omega_{\mathrm{DM}}$ the initially assumed dark matter density. Unlike Ref.~\cite{2019PhRvD..99l1302V}, $a_\star$ is set to the early Universe, well before matter domination and therefore for late decays  ($\Gamma t_\star \ll 1$), such as what we consider here, the effects of decays in the early universe are negligible. The quantities $a_D$ and $H_D$ are the scale factor and the corresponding Hubble parameter at scale factor $a_D$ (and time $t_D = t(a_D))$ of decaying particles. The physical picture behind this expression is that due to conservation of momentum, the massive daughter is produced with a non-zero velocity that later redshifts away as the Universe expands and the particles cool down.

\begin{figure*}
\centering
\includegraphics[width=0.9\columnwidth]{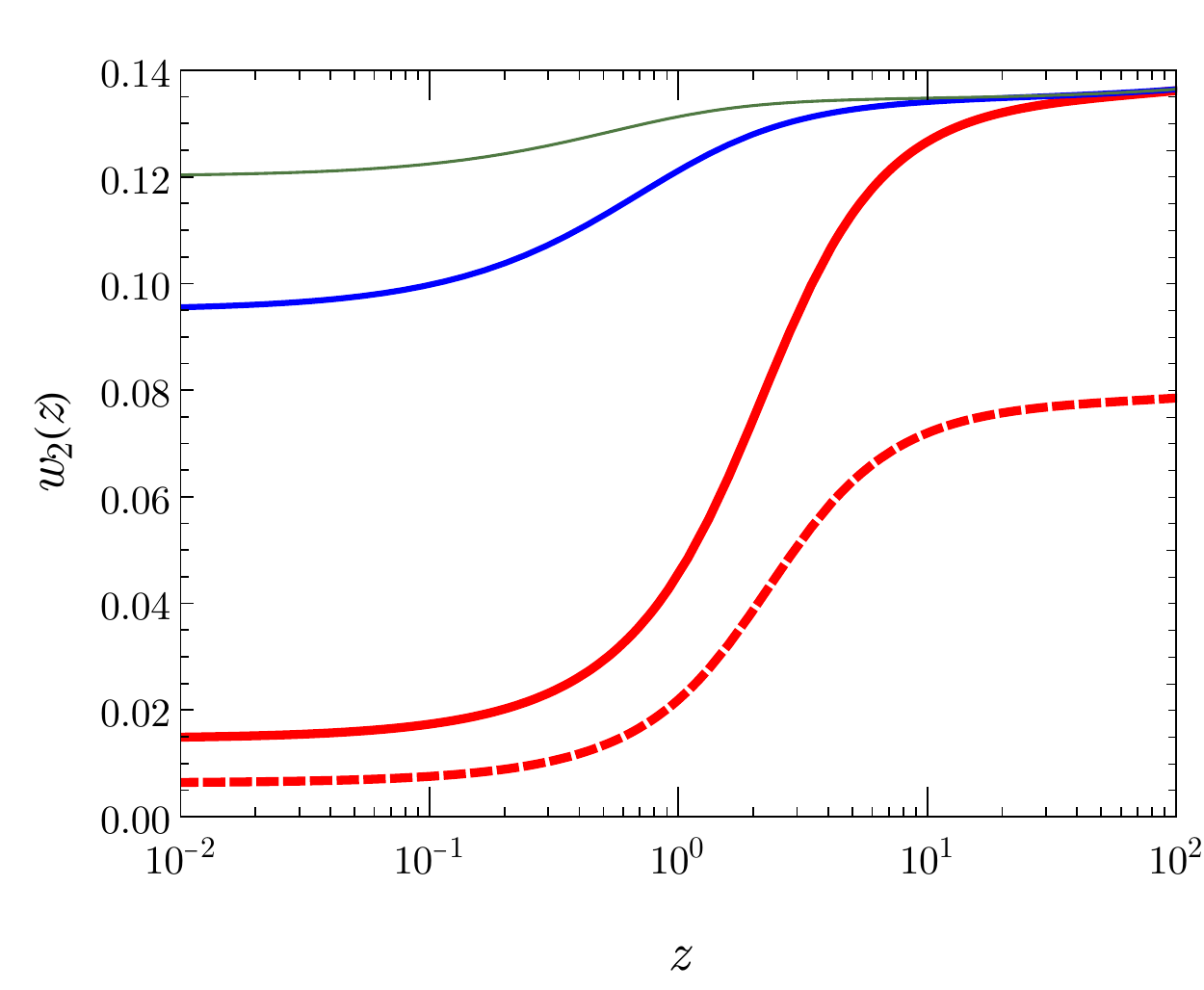}
\includegraphics[width=0.9\columnwidth]{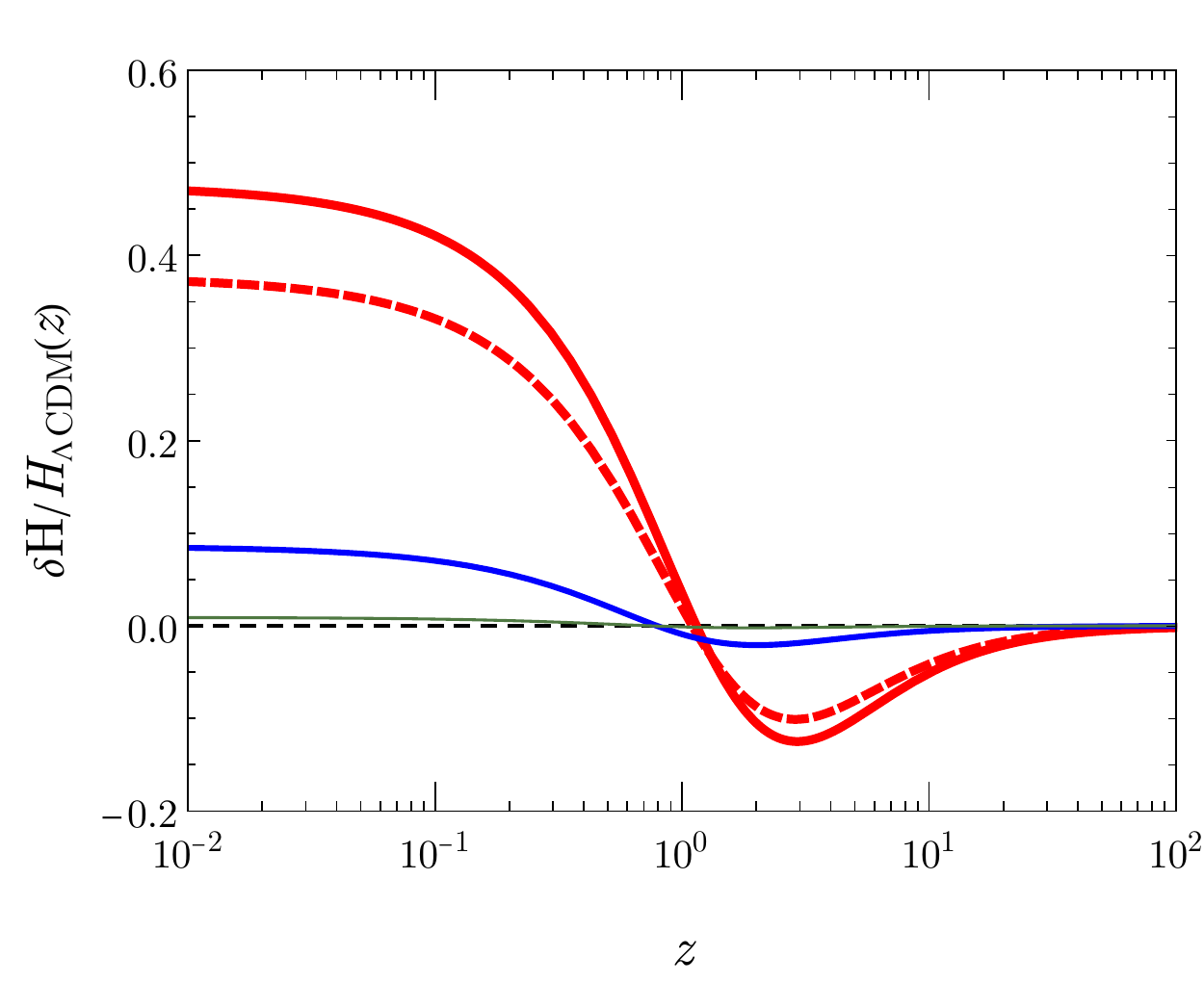}
\caption{{\it Left}: Evolution of $w_2$ for $\{\Gamma / { \mathrm{Gyr}}^{-1},\epsilon\} = \{0.01,0.45\}$, $\{0.1,0.45\}$, $\{1,0.45\}$ and $\{1,0.4\}$, shown as a thin green, medium blue,  thick red and thick dashed red lines respectively.  The decay width $\Gamma$ controls the time and the amount that $w_2$ changes state while $\epsilon$ controls the absolute magnitude by correlation to the initial velocity of the massive decay products -- see text for discussion. {\it Right}: A consequence of the introduction of the two-body dark matter decay model is the effect on the expansion rate of the Universe \cite{2019PhRvD..99l1302V}.  The right panel shows the ratio of the expansion rate in the presence of decays over a baseline $\Lambda$CDM Universe as described at the beginning of \Sec{sec:decaying_dm} for the same three models as figure to the left. The most important features is an increase of the expansion rate of the Universe compared to $\Lambda$CDM at late times while we see a smaller in magnitude decrease on earlier times.  }
\label{fig:fig1}
\end{figure*}

A key feature that distinguishes this model from other decay scenarios is the dynamical properties of the massive daughter particle $\chi$'s equation of state, $w_2$. The left panel of  \Fig{fig:fig1} shows  the equation of state for four different sets of lifetimes of particle decays and the parameter $\epsilon$. Particles at creation are behaving as warm dark matter, with non-zero equation of state, that ``slow-down" as the universe expands. The  initial amplitude of $w_2$ is determined by the value of  $\epsilon$: the velocity of the particle at decay is $v_2 \sim \beta_2$, and as $w_2 \sim \beta_2^2 / 3 \sim \epsilon^2 / 3(1- \epsilon)^2$. As $\epsilon$  takes values between 0 and 1/2, we see that the range of values of $w_2$ at decay is between 0 and 1/3 -- see \cite{2014PhRvD..90j3527B} for more details. 

At any given time, the equation of state of all daughter particles is collectively encapsulated by $w_2$; For example, if one were to calculate the equation of state today, the aforementioned determination of $w_2$ includes all particles that decayed in the past (and whose velocity has been redshifted, i.e., slowed down) as well as particles that are decaying currently. The weight of each population (from the past to the present) is completely determined by the decay width $\Gamma$ which governs the input rate of new particles with a given speed in the dark matter fluid. At small values,  of  $\Gamma$ the injection of new particles is sustained for longer and the equation of state remains constant regardless of the initial speed. Conversely, at larger values of $\Gamma$ most of the massive daughter particles are produced early on and their speeds have more time to redshift away to small values (unless of course the particles are born non-relativistic). An additional subtle consequence of varying $\Gamma$ is that it controls $\dot{w}_2$, i.e., the time derivative of the equation of state. For example if $\Gamma$ is of order the inverse of the matter--dark energy equality timescale then  $\dot{w}_2$ is larger compared to a $\Gamma$ that is much smaller.

\subsection{Effect of decays on $H(z)$}\label{sec:Hz} 
A very important consequence of the introduction of the dark matter decay model is the effect on the expansion rate of the Universe as decays can change the relative amount of relativistic and non-relativistic components that enter in the calculation of the Hubble parameter as a function of redshift \cite{2019PhRvD..99l1302V},     
\begin{equation}
H^2(a) \equiv \left(   \frac{\dot{a}}{a}  \right)^2 = \frac{8 \pi G}{3}  \sum_i \rho_i(a) ,  
\label{eq:Ha}
\end{equation} 
where 
\begin{eqnarray} 
\sum_i \rho_i(a) &=& \rho_0(a) + \rho_1(a) + \rho_2(a) \nonumber \\
&+& \rho_r(a) + \rho_\nu(a) + \rho_b(a) + \rho_\Lambda. 
\label{eq:friedmann} 
\end{eqnarray} 
Here, $\rho_0$, $\rho_1$ and $\rho_2$ correspond to the energy densities of the parent dark matter particle, and the massless and massive daughters respectively, and $\rho_r$, $\rho_\nu$, $\rho_b$ and $\rho_\Lambda$ are the energy densities of photons, neutrinos, baryons and dark energy respectively. Note that in the decaying dark matter case we study here,  all dark matter densities ($\rho_0$, $\rho_1$ and $\rho_2$) in \Eq{eq:friedmann} are not only scale factor dependent but also depend on time according to  Eqs.~(\ref{eq:rho0}--\ref{eq:rho2}).

The right panel of \Fig{fig:fig1} shows  the ratio of the expansion rate in the presence of decays over a baseline $\Lambda$CDM Universe as described at the beginning of \Sec{sec:decaying_dm}. Qualitatively, decays manifest themselves in the value of the expansion rate as a decrement at redshifts $z \gtrsim 1$ and as an increment at redshifts $z \lesssim 1$. 

The initial deceleration at redshifts $z \gtrsim 1$ is caused because during matter domination a fraction of dark matter (the exact amount governed by $\Gamma$ and $\epsilon$) transitions to radiation, with energy density evolution governed by \Eq{eq:rho1}. The pressure due to radiation transfer effectively acts as a break to the expansion rate. This effect explains why larger values of $\epsilon$ as well as higher decay rates $\Gamma$ cause a larger dip; the higher the values the larger amount of energy is transferred between the two species.

This transfer of energy into radiation is also the same reason we observe an acceleration in later times. As matter is depleted into radiation the matter-dark energy equality is shifted to earlier redshifts, allowing for higher value of $H_0$ at late times. As before, larger values of $\epsilon$ and $\Gamma$ cause a more dramatic effect as the decays become more effective during the lifetime of the Universe. This very characteristic makes this model a promising candidate to solve the $H_0$ tension by matching the extrapolated value from early Universe estimations to the late Universe measurements as was shown in \cite{2019PhRvD..99l1302V}.

\subsection{Effect of decays on the matter power spectrum}\label{sec:Growth} 
Measurements of the growth of structure provide a wealth of information regarding the abundance and properties of dark matter and dark energy and are complimentary to distance measurements such as baryon acoustic oscillations and supernovae. The time-dependence of the growth of structure using the matter power spectrum is sensitive to the temporal evolution of dark matter and as such current (e.g., DES, eBOSS) \cite{2020arXiv200211124D, 2020MNRAS.492.4189I} and future experiments (LSST, PFS, Euclid and W$FIRST$)~\cite{2009arXiv0912.0201L, 2018SPIE10702E..1CT, 2011arXiv1110.3193L, 2012arXiv1208.4012G} are able to constrain properties of dark matter, modifications to gravity as well as the time-dependence of dark energy. 

We can quantify the effects of dark matter decays on the growth factor in the following way. Given a scale invariant power spectrum, the growth of linear matter fluctuations (defined as $\delta = \delta \rho / \rho \ll 1$) is governed by 
\begin{equation} 
\ddot{\delta}  + 2 H \dot{\delta} - 4 \pi G \rho_M \delta = 0, 
\label{eq:growth}
\end{equation} 
where the derivatives are with respect to time, $G$ is Newton's constant, and $\rho_M$ is the matter density, where in the case of decaying dark matter it is given by
\begin{equation} 
\rho_M=\rho_0+(1-3 w_2)\rho_2+\rho_b, 
\label{eq:rhoM}
\end{equation} 
where $w_2$ is given by \Eq{equ:w_2}. 
The solution of \Eq{eq:growth} provides the growth factor, defined as $D(a) = \delta(a) / \delta(a=1)$, normalized to unity today ($a=1$). A change in the time evolution of $\rho_M$ in equation \Eq{eq:growth} changes both, the second and third terms, and it is the competition between these two terms that sets the net effect of dark matter decays on the growth factor. 

\begin{figure}[t]
\centering
\includegraphics[width=0.9\columnwidth]{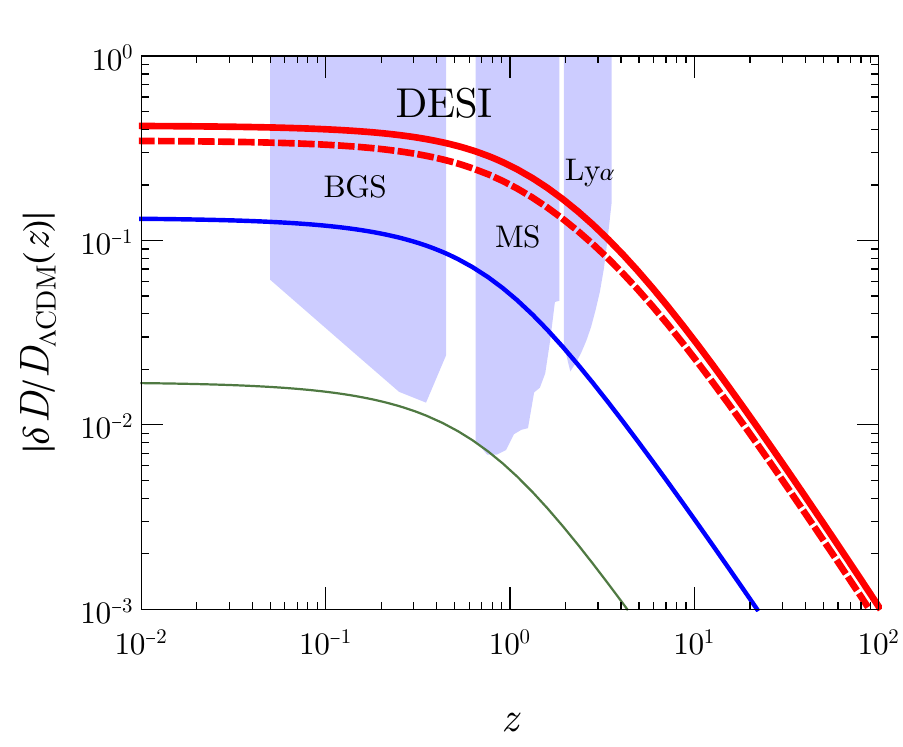}
\caption{Evolution of the fractional linear growth factor D(z) change in the presence of two-body decaying dark matter over the same  baseline $\Lambda$CDM model and same four models as \Fig{fig:fig1}. At late times the decaying dark matter scenario erases structure which for part of the parameter space of the model can be probed by the Dark Energy Spectroscopic Instrument (DESI) \cite{2016arXiv161100036D}.}
\label{fig:fig2}
\end{figure}

Figure~\ref{fig:fig2} shows the fractional deviations in the growth factor in the presence of dark matter decays compared to a fiducial $\Lambda$CDM model as described above for the four examples of decaying dark matter scenarios discussed earlier.  At early times ($z \gg 1$), before dark matter decays become important (i.e., $\rho_1 \ll \rho_0$), the universe is matter dominated, and therefore $\delta(a) \sim a$ as in the case of $\Lambda$CDM. As decays take over, the dark energy -- matter equality is reached earlier than in $\Lambda$CDM, and the growth of structure is suppressed. This effect is more prominent at late times ($z \lesssim 1$).   As with the evolution of the expansion rate, $H(z)$, larger values of $\Gamma$ and $\epsilon$ cause larger deviation from $\Lambda$CDM as more matter energy budget is transformed into radiation, a non-surprising result since the growth factor itself is obtained (partly) by the integral of $H(z)$. 

An additional way to characterize the linear growth of structure is by looking at the matter power spectrum $P(k)$, on a scale $k$, defined as the 2-point correlation function of over densities $\langle \delta_{\bf{k}} \delta_{\bf{k}}\prime \rangle = (2 \pi)^3 \delta ({\bf{k}} - {\bf{k}}\prime ) P(k)$. The variance of mass fluctuations at a physical scale $R$ is then obtained from
\begin{equation} 
\sigma_R(a) = \frac{1}{2 \pi} \int_0^\infty k^3 P(k,a) \tilde{W}^2_R(k) \, \mathrm{d}\ln k, \label{eq:sigma8}
\end{equation} 
where $\tilde{W}_R$ is the Fourier transform of the top-hat window function. \Eq{eq:sigma8} gives the variance of the amplitude of fluctuations over a sphere with radius $R$ at scale factor $a = 1 / ( 1 + z)$. It is customary to quote a value of the variance over a sphere of radius $8 h^{-1} {\rm{Mpc}}$, commonly referred to as $\sigma_8$.

The amplitude of the power spectrum around~$k \sim 1 \, h \, {\rm Mpc}^{-1}$ (the scale probed by \si8) characterizes rare events on the tail of the distribution and as such it is very sensitive on the cosmological parameters. Observationally it is accessible through measurements of the galaxy cluster mass function in optical surveys \cite{2020arXiv200211124D} as well as through the Sunyaev-Zel'dovich effect on the CMB \cite{2018arXiv180706209P}. Introducing decaying dark matter has the effect of a time dependence decrease on the matter density (\Eq{eq:rhoM}) and a resulting increase in $\rho_1$ that behaves as radiation. The result is a reduction in the value of \si8 \cite{2015JCAP...07..009C}. One of the reasons that late universe dark matter decays received interest in explaining the $H_0$ tension is due to this side-effect -- the reduction of \si8 and thus reducing also the tension that exists between the values of \si8 deduced by CMB measurements and cluster number counts from optical surveys \cite{2019PhRvD..99l1302V}. 

\begin{figure*}
\centering
\begin{tabular}{c}
\includegraphics[width=0.9\columnwidth]{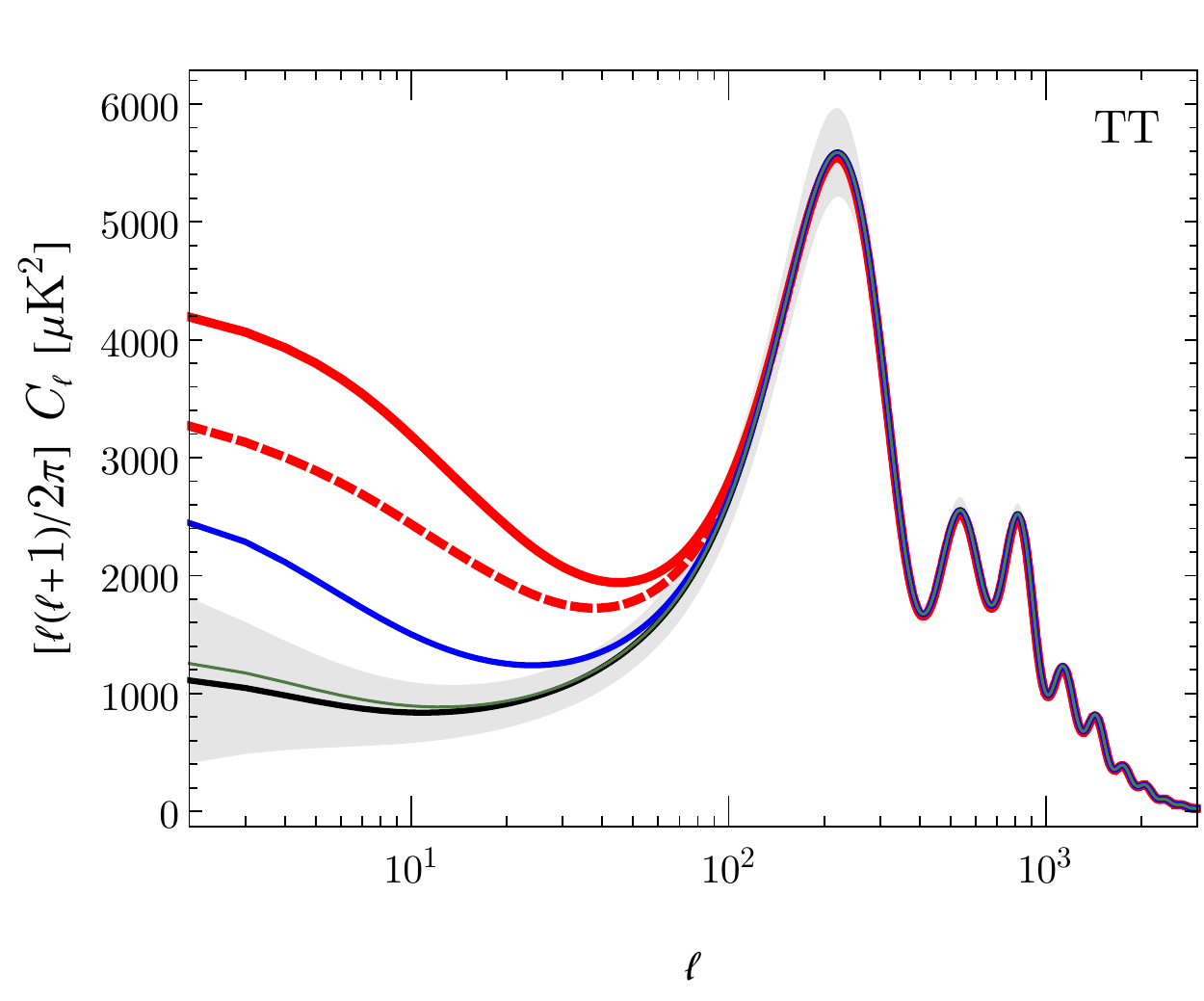}  
\includegraphics[width=0.9\columnwidth]{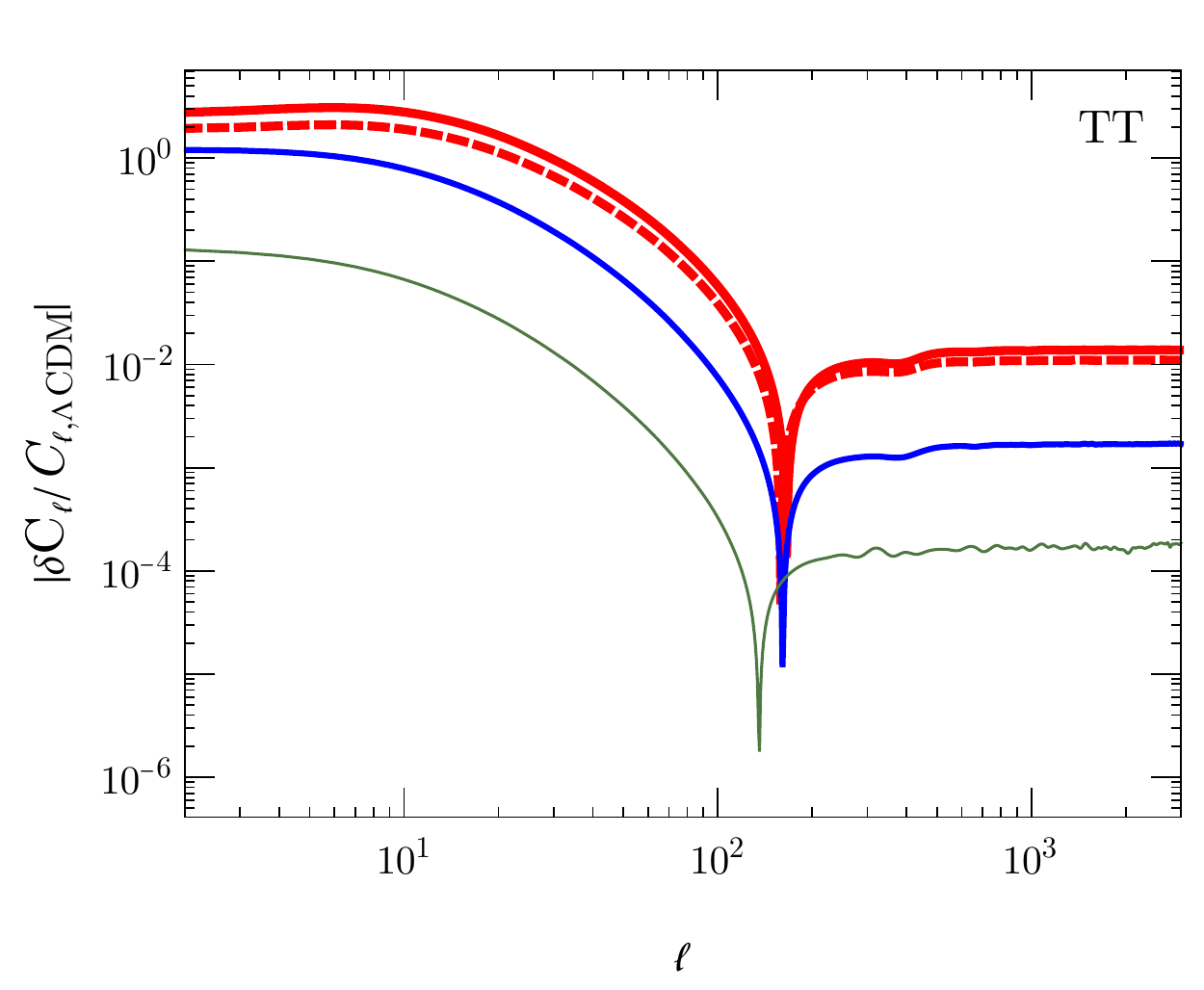} \\
\includegraphics[width=0.9\columnwidth]{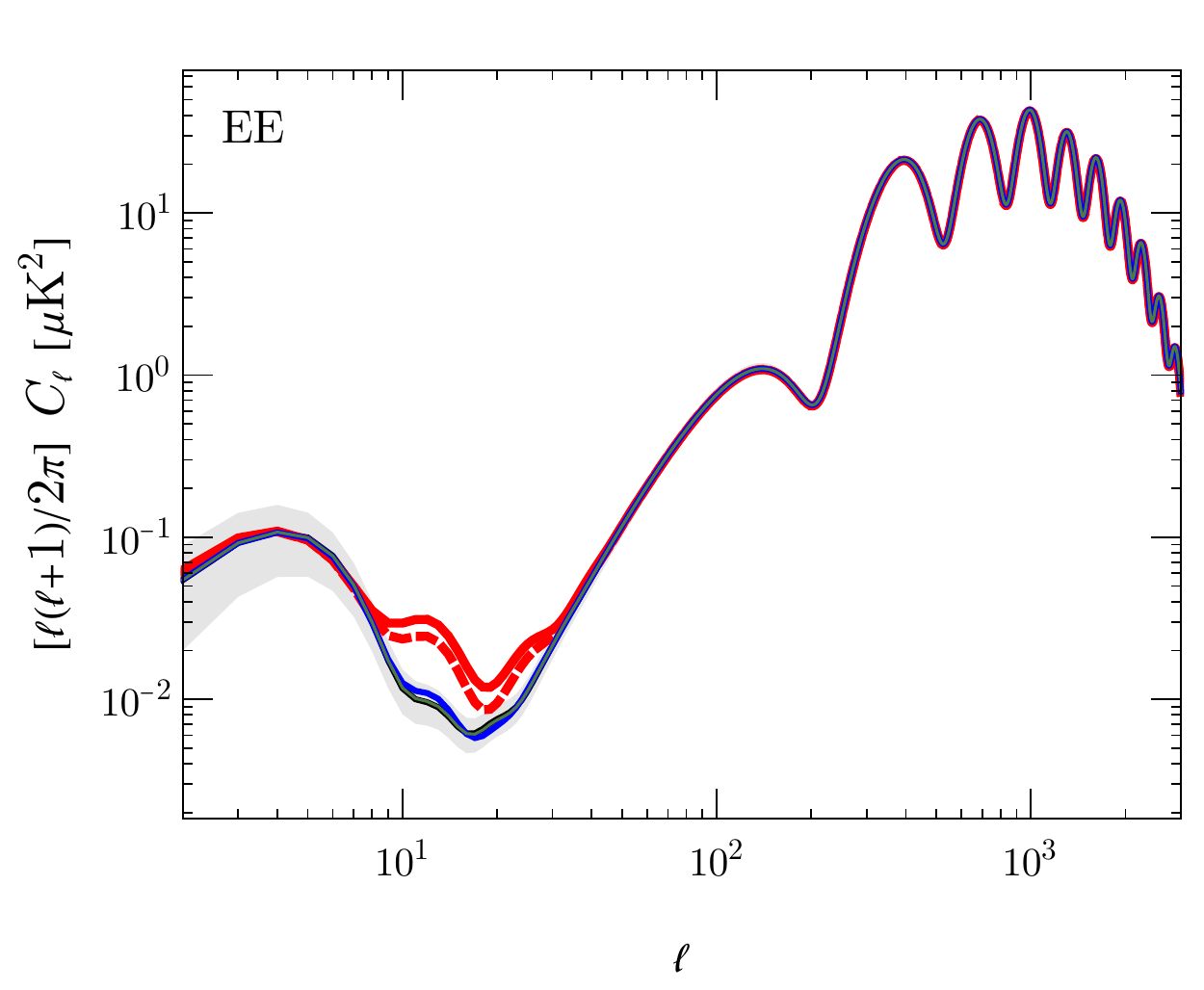}  
\includegraphics[width=0.9\columnwidth]{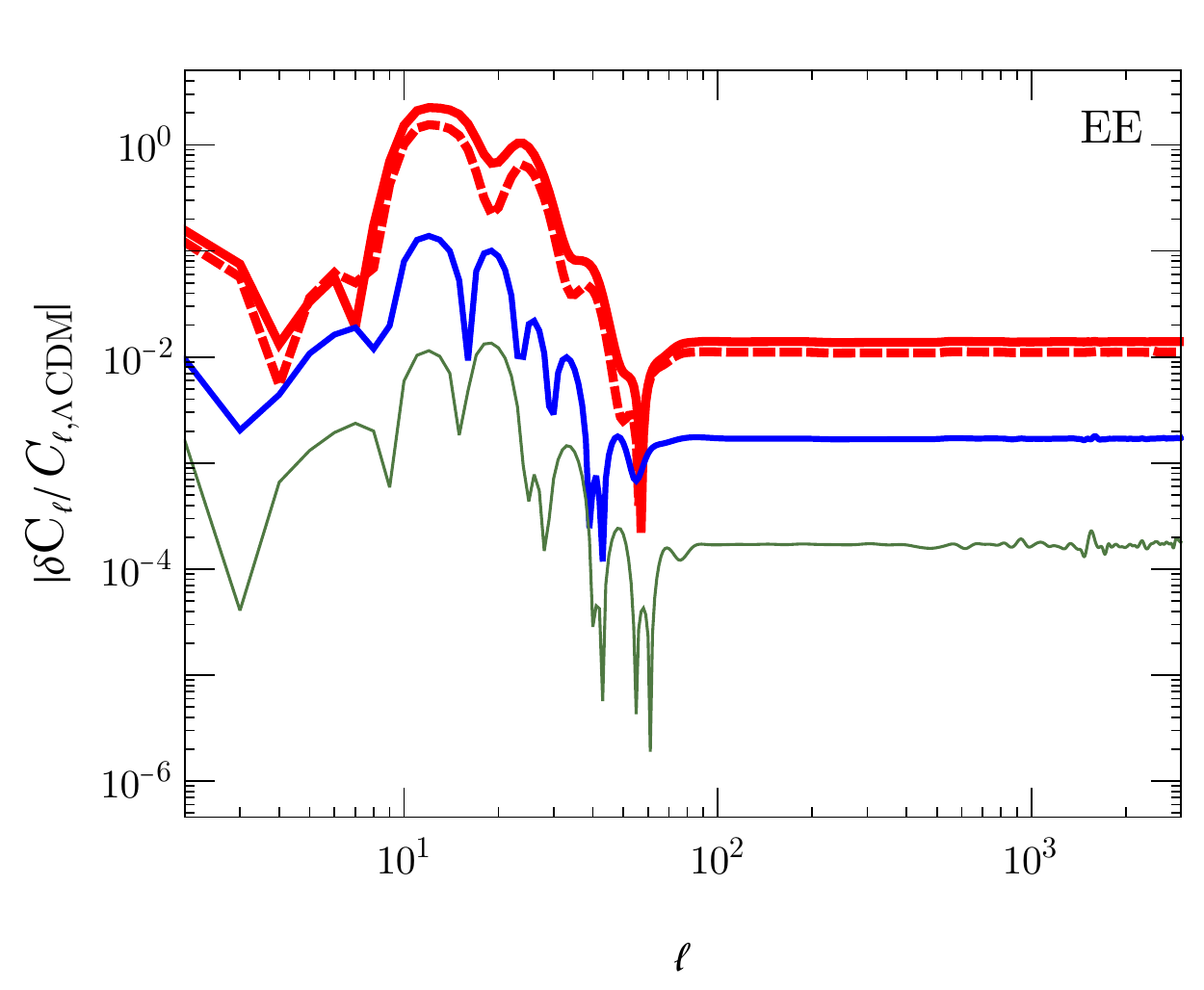} 
\end{tabular}
\caption{The cosmic microwave background TT (top) and EE (bottom) power spectra (left) and fractional changes (right) for the four examples of decaying dark matter as in \Fig{fig:fig1}. Dark matter decays appear in the CMB power spectrum as increased amplitude  in the late-ISW effect as the dark matter -- dark energy equality is moved to earlier times. Additionally, the altered expansion history results in additional scatterings during reionization, producing larger polarization correlations at low multipoles.}
\label{fig:fig3}
\end{figure*}

\subsection{Effect of decays on CMB anisotropies} \label{sec:CMB}

The CMB power spectrum is an imprint of the conditions of the universe at the epoch of recombination as well as an encoder of all processes that can alter that spectrum between recombination and today. The power spectrum is written as a linear combination of physical processes. These include (among others) the intrinsic photon temperature fluctuations at recombination, fluctuations due to perturbations in the gravitational potential (known the Sachs-Wolfe effect), a Doppler effect due to the acoustic motion of the photon-baryon fluid as well as the relative motion between the observer and the last scattering surface, and temperature anisotropies that arises from the time-dependent gravitational potential integrated along the line of sight, known as the Integrated Sachs--Wolfe (ISW) effect.~\cite{2002ARA&A..40..171H}

The ISW effect is present if there is residual radiation during recombination, in which case potentials inside the horizon can decay. This is known as the early-ISW effect, and leaves an imprint on scales smaller than the horizon at recombination. The late-ISW effect arises as light travels from the last scattering surface, through time-dependent potentials. It appears once the universe is no longer matter dominated (i.e., dark energy begins to dominate),  and manifests itself on large angular scales due to the proximity of the origin of the effect and the large horizon size at dark energy domination. 

In the absence of anisotropic stresses the scalar temperature anisotropy $\Theta_{\rm{ISW}}(\hat{\bf{p}})$ due to the ISW effect along a direction $\hat{\bf{p}}$ is 
\begin{equation} 
\Theta_{\rm{ISW}}(\hat{\bf{p}}) \equiv \frac{\Delta T^{(s)} ( \hat{\bf{p}})}{T} \sim \int \frac{\partial \Phi}{\partial \eta}  \, \mathrm{d} \eta.
\label{eq:ThetaISW}
\end{equation} 
Here,  $\Phi$ is the Newtonian potential perturbation to the metric, the derivative is with respect to conformal time $\eta$, and the limits of integration are from recombination to the present time.  Expansion of \Eq{eq:ThetaISW} in spherical harmonic gives the anisotropy in a direction $\hat{\bf{p}}$ as 
\begin{equation} 
\Theta(\hat{\bf{p}}) \sim\sum_{\ell=0}^\infty \sum_{m=-\ell}^{\ell} i^\ell \int  \frac{\partial \Phi}{\partial \eta}  j_{\ell m}(k \eta) Y_{\ell m}( \hat{\bf{p}}) Y_{\ell m}^*( \hat{\bf{k}}) \mathrm{d} \eta  \mathrm{d}k, 
\end{equation}
where $j_{\ell m}(k \eta)$ is the spherical Bessel function. The integral is evaluated from the last scattering surface to the present. Most of the Bessel function contribution to the integrant  comes from scales that are of order $\ell \sim k \eta$, which means as $\eta$ increases larger scales become more dominant. This explains the fact that the late-ISW effect appears on large scales compared to the horizon size at recombination. 

Using the orthogonality of spherical harmonics and the potential as given by Poisson's equation, we can get the angular power spectrum in the case of decaying dark matter, 
\begin{eqnarray} 
C_{\ell} &=& \langle \Theta_{\ell m}\Theta_{\ell m}^* \rangle \nonumber \\
&\sim&  \int \mathrm{d}k P(k) \left[ \int \mathrm{d} \eta \, a^2 \left( \frac{\partial \lambda}{\partial \eta} + 2  \mathcal{H} \lambda \right) \right]^2. 
\label{eq:Cleq}
\end{eqnarray} 
Here, $P(k)$ is the matter power spectrum (assumed of the form $P(k) \sim k^n$, with $n\approx -1$), $\lambda = \rho_M \delta$, and $\mathcal{H}$ is the conformal Hubble parameter. 

In the limit where decays are not present, i.e., the matter density scales as $\rho_M \sim a^{-3}$, the term in the parentheses in \Eq{eq:Cleq} reduces to the known result  $\sim ( f - 1) D \mathcal{H} j_\ell (k \eta)$, where $D$ is the growth factor, and $f \equiv  \mathrm{d} \ln D / \mathrm{d} \ln a$ (note that in principle \Eq{eq:Cleq} contains a factor of $e^{-\tau}$, where $\tau$ is the optical depth to the last scattering surface, which we assume here to be $e^{- \tau} \approx 1$ for late-universe decays). 

The qualitative effects on decaying dark matter on the power spectrum can be understood in the following way. Increasing the decay width of dark matter $\Gamma$, pushes the dark-energy -- matter radiation to higher redshifts, i.e., earlier times, thus power is increased on low-$\ell$ scales, with progressively larger values of $\ell$ affected.  An increase in $\epsilon$ corresponds to  a larger branching ratio to radiation and a massive daughter particle with increasing initial velocity. Both of these effects result  in an earlier shift to dark energy domination and again an increase of power on low-$\ell$'s. The top two panels of \Fig{fig:fig3} shows these effects on the temperature power spectrum for the four representative cases we use as an example. 

In addition to the scalar temperature fluctuations, decaying dark matter also leaves an imprint on the polarization of the CMB~\cite{2002ARA&A..40..171H}. The CMB polarization is produced as photons experience Thompson scattering off free electrons. The majority of these interactions occur near the surface of last scattering, and produce the large anisotropic peaks above $\ell>100$. Since these interactions occur around recombination, dark matter has yet to decay (in late-universe decays) and the process is identical to the $\Lambda$CDM model for typical decay properties.

However, Thompson scattering can also take place much later during the epoch of reionization. The cumulative level of scattering interactions is typically quantified  by the integrated reionization optical depth~\cite{2018arXiv180706209P}, 
\begin{equation}
\tau = n_{\rm H}(0)\,c\,\sigma_{\rm T} \int_0^{z_{\rm max}} { d}z\,x_e(z)\frac{(1+z)^2}{H(z)}, 
\label{eq:tau}
\end{equation}
where $x_e(z)=n_e^{\rm reion}(z)/n_{\rm H}(z)$ is the free electron fraction, $n_e^{\rm reion}(z)$ as the free electron number density, and $n_{\rm H}(z)$ is the hydrogen nuclei number density. The quantities $\sigma_{\rm T}$ and $c$ are the Thompson scattering cross-section and the speed of light respectively. As noted in Ref.~\cite{2018arXiv180706209P}, reionization results in a suppression factor of order $e^{-2\tau}$ to the anisotropies above $\ell \approx 10$.

In addition to the expected suppression on large scales, Thompson scattering at reionization creates polarization anisotropies for $\ell \leq 30$ that appear as a bump in the polarization spectra at low $\ell$. The height of the bump is proportional to $\tau^2$, and corresponds to a scale comparable to the Hubble horizon during the epoch of reionization. Ref.~\cite{2018arXiv180706209P} finds that the CMB $\tau$ constraint (and thus the height of the bump) is fairly model-independent from the free electron fraction, $x_e$; however, the shape of the bump depends  on $x_e$~\cite{1997ApJ...488....1Z,2003ApJ...583...24K}. Note that there is a degeneracy between  $x_e$ and $H(z)$ (as is evident in \Eq{eq:tau}). Late universe decaying dark matter changes the expansion rate (see \Fig{fig:fig1}). A reduced expansion rate implies an increased optical depth for the same ionization fraction. This implies reionization occurs over a longer period of time, and therefore scattering interactions increases thus changing the shape of the bump of the polarization anisotropies between $10<l<30$.

All discussions up to this point have assumed no large scale structure lensing (i.e., no lensing effects were included in \Fig{fig:fig3}. As the CMB photons traverse the Universe, they are gravitationally lensed by foreground structure resulting in multiple effects. Here, we focus on the smoothing of  small scale anisotropic peaks and troughs as a result of a convolution between the un-lensed spectra and the lensing potential of large scale  structure~\cite{2006PhR...429....1L}: the strength of the lensing potential is directly tied with the amount of structure present. As the amplitude of lensing effects are directly related to the amount of intervening structure (and its growth) along the line of sight, we expect that as decaying dark matter suppresses the growth of structure in the Universe, it will in turn reduce the amount of expected lensing. 

\begin{figure}[t]
\centering
\includegraphics[width=0.9\columnwidth]{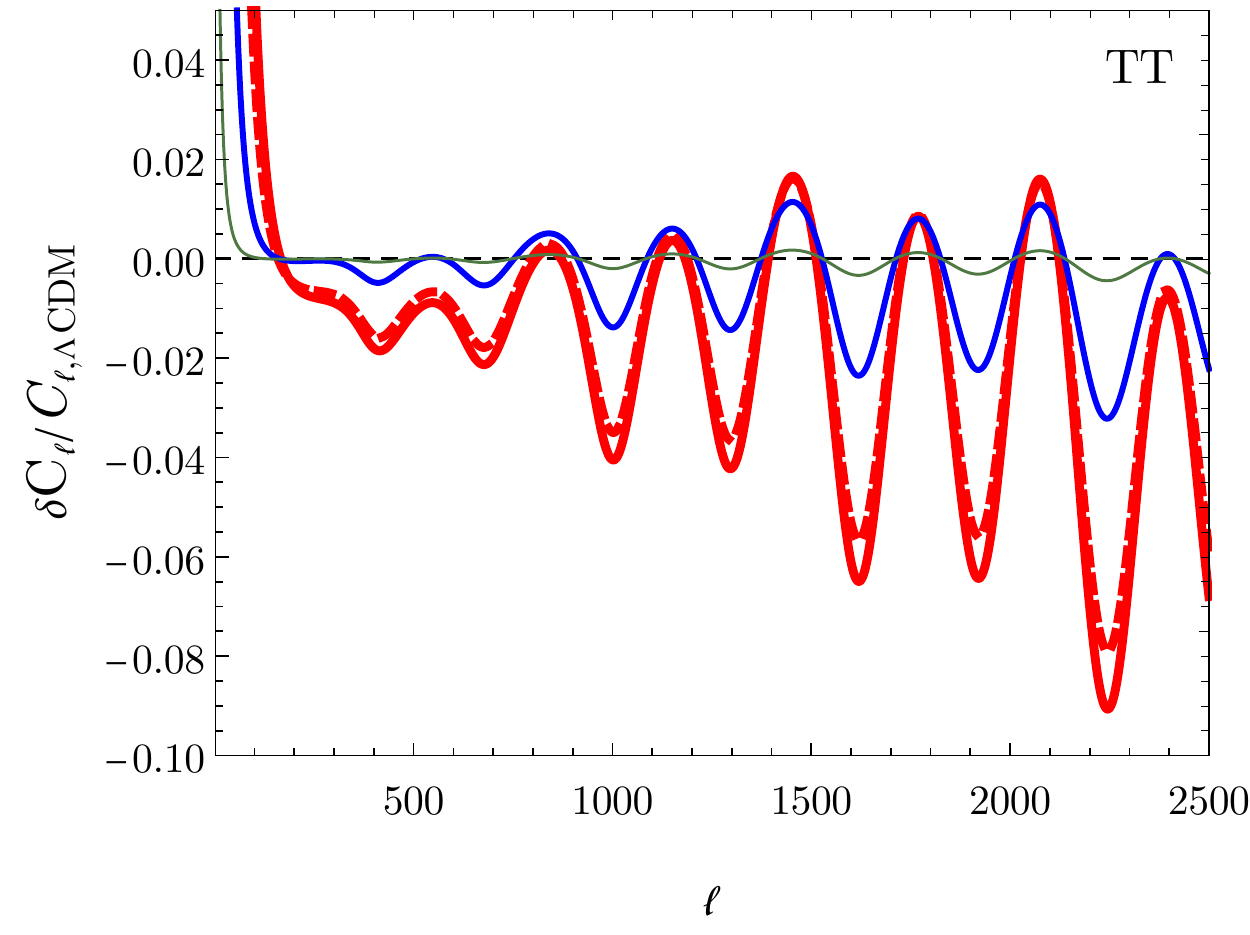} 
\caption{The fractional change in the lensed CMB TT power spectrum for the four cases of decaying dark matter as \Fig{fig:fig1}. Lensing induces larger variations from $\Lambda$CDM for decaying dark matter as reduction in the growth of structure reduces the amount of lensing at small scales.}
\label{fig:fig4}
\end{figure}

In \Fig{fig:fig4} we show the percent change of the high multipoles of {\it{lensed}}  TT correlations on the CMB in the decaying dark matter scenario as compared to the lensed fiducial cosmological model (see the beginning of \Sec{sec:decaying_dm}). As expected, reducing the growth of structure (due to decays) leads to  a reduced lensing effect on the power spectrum. The deviations from $\Lambda$CDM's lensing spectra are at percent level for even the most minimal decay parameters  outpacing the deviation from the un-lensed spectra alone as compared with \Fig{fig:fig1}. The oscillations in the lensed ratio correspond directly with the oscillations in the un-lensed spectra, with peak differences in the lensed case matching troughs in the un-lensed one. 

In summary, the overall effect of decaying dark matter on  CMB anisotropies is an increase in both temperature and polarization at low $\ell$ and an increase in the magnitude of oscillations at high $\ell$. These variations at low and high $\ell$ are due to changes in the expansion rate at late times and a decease in the lensing potential, respectively. In the next section we will use these two physical effects to constrain the two-body decaying dark matter scenario. 

\section{Constraints} \label{sec:constraints}
We can constrain the decay width $\Gamma$, and fraction of rest mass energy that goes to radiation, $\epsilon$, by performing a Markov Chain Monte Carlo fit on the decaying dark model of \Sec{sec:decaying_dm} using \textsc{MontePython}~\cite{2018arXiv180407261B} and the Planck 2018 TTTEEE+low$l$+lowP+lensing data sets as well as BAO (SDSS DR7\cite{2010MNRAS.404...60R}, 6FD\cite{2011MNRAS.416.3017B}, MGS\cite{2015MNRAS.449..835R}, BOSS DR12\cite{2017MNRAS.470.2617A}, eBOSS Ly-$\alpha$ combined correlations\cite{2019A&A...629A..85D,2019A&A...629A..86B}) and the Pantheon SNIa catalog\cite{2018ApJ...859..101S}. We choose these data sets as they include all the combinations of measurements from Planck including lensing which potentially can help constraint the effects of the decaying dark matter on the structure formation. We assume the same $x_e$ history for all models and vary  $\Lambda$CDM as directed in Ref.~\cite{2018arXiv180706209P} with the addition of the two variables, $\Gamma$ and $\epsilon$.

\begin{figure*}[ht]
\centering
\includegraphics[width=2.0\columnwidth]{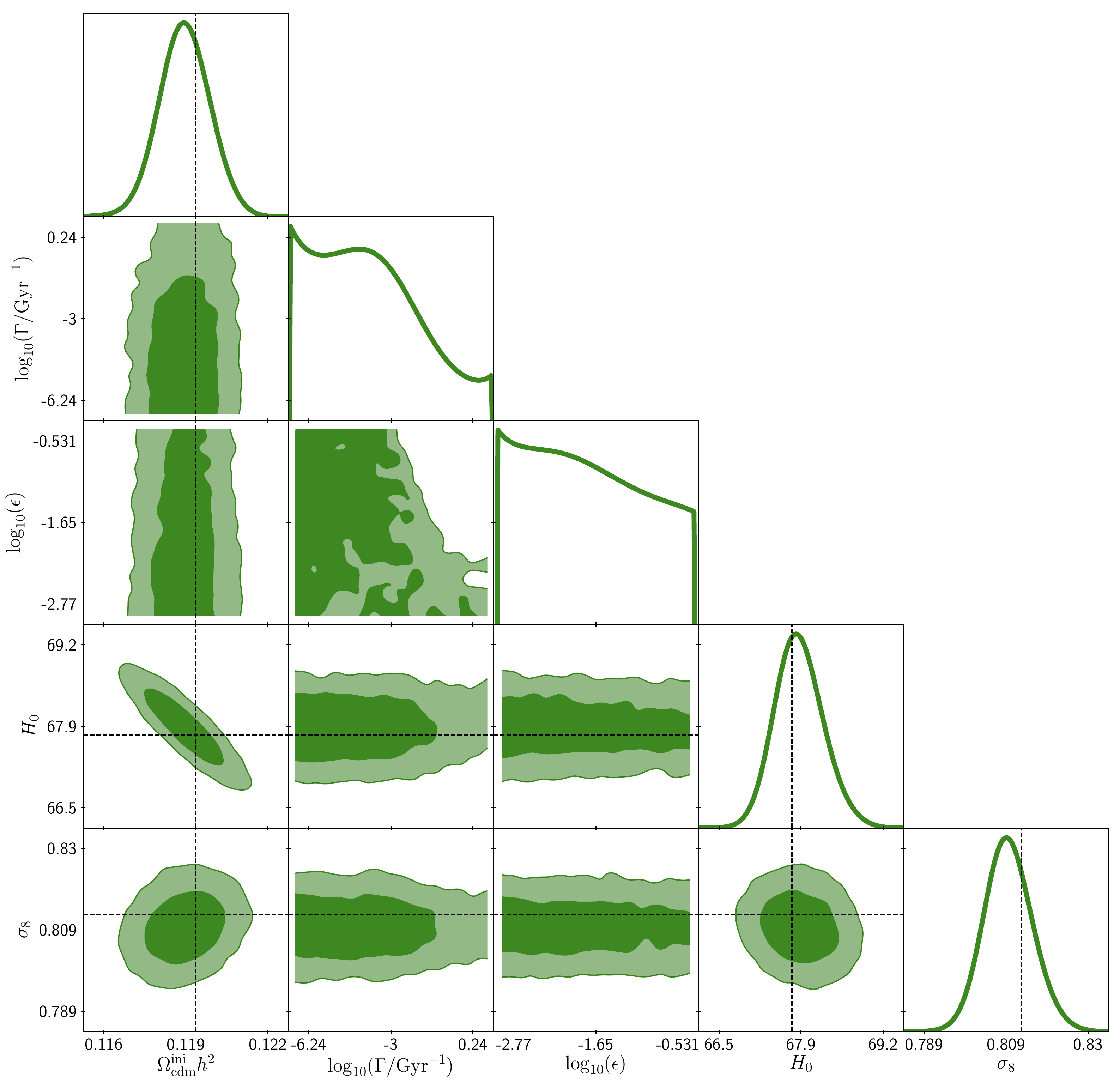}
\caption{The 2-d contour plot for a subset of parameters in decaying dark matter fit to Planck 2018 TTTEEE+low$l$+lowP+lensing+BAO+Pantheon. The preferred region for all $\Lambda$CDM parameters are the same as $\Lambda$CDM. The preferred region in the $\Gamma$-$\epsilon$ contour corresponds to a region where the effects from decays are minimal.}
\label{fig:fig5}
\end{figure*}

We calculate the resulting cosmology and CMB anisotropies  with a modified version of \textsc{CLASS}\footnote{\url{http://class-code.net/}}
~\cite{2011JCAP...07..034B}.  We calculate the present-day dark matter density by the shooting method as described in Ref.~\cite{2014JCAP...12..028A}. For computational convenience we follow Ref.~\cite{2014PhRvD..90j3527B} for the implementation of \Eq{equ:w_2} in \textsc{CLASS}. More specifically, we can write the equation of state  as
\begin{equation}
w_2(t) = \frac{1}{3} \langle v_2^2(a) \rangle, 
\end{equation}
where $v_2$ is the speed of the massive daughter particle at scale factor $a$, that was produced earlier when $a = a_D$. With $\tilde{a} \equiv a_D/a$, the average speed of the massive daughter can be written as 
\begin{equation}
\langle v^2(t) \rangle = \int_{t_\star}^t v^2(\tilde{a}) \dot{n}_2{\rm d}t_D \Big/ \int_{t_\star}^t \dot{n}_2{\rm d}t_D , 
\end{equation}
where
\begin{equation}
v_2^2(\tilde{a}) = \frac{\tilde{a}^2 \beta^2_2}{1+\beta_2^2 \left[\tilde{a}^2-1\right]},
\end{equation}
and $\dot{n}_2 \equiv \mathrm{d}n_2/\mathrm{d}t_D$ is the time derivative of the massive daughter's number density. The latter is obtained by setting $\dot{n}_2 = -\dot{n}_0$ as for every decayed parent particle there is one massive particle created and can be written as $\dot{n}_2= \Gamma \rho_0(t_D)/m_0 \tilde{a}^3$ where the factor $\tilde{a}^3$ scales the number density to its value at the time the velocity is calculated. We use this formulation in order to assists in calculations during early time steps when $\psi$ decays are negligible.

With the background evolution defined, we can now turn our attention on the perturbations. The treatment of the massive daughter source terms involved in perturbing the evolution of the universe is non-trivial as it is possible for it to be warm at production, while particles that were produced earlier may have already ``cooled down". In order to make the computations in CLASS efficient we treat the massive daughter particle contributions with the assumption they can be separated into a relativistic and non-relativistic component, termed hot and cold respectively,  each characterized  with an equation of state
\begin{equation}
\rho_{2, \, {\rm hot}} = 3 \, w_2 \, \rho_2, \qquad \rho_{2, \, {\rm cold}} = (1-3 \, w_2) \, \rho_2.
\label{eq:rho2w}
\end{equation}
An important note to make here is that the massive daughter particle density $\rho_2$ and its equation of state $w_2$ refer to their background evolution values calculated as described earlier and thus don't take any perturbations into account. While this approximation may break down for ``warm-like" cases, it is exact for $\epsilon=0$ and $0.5$ as initial velocities are  $0$ and $c$ respectively. As mentioned earlier, the equation of state {\it at decay} scales as $w_2 (a_D) \sim \epsilon^2 / 3 (1 - \epsilon)^2$, however the equation of state of a population after a considerable amount of decays (i.e., timescales of order the lifetime of the particle) will always have $w_2 (a) < w_2 (a_D)$ as the particle momentum redshifts due to the expansion of the universe. With this we can justify our assumption in \Eq{eq:rho2w} because as $t \gg \Gamma^{-1}$, $w_2 \rightarrow 0$ and the  massive daughter asymptotically behaves as cold dark matter while based on the results of \cite{2019PhRvD..99l1302V} we shouldn't expect very warm daughter particles to be favourable.

In \Fig{fig:fig5} we show the 2-D contours of the posterior distribution of the free and derived parameters as determined by the MCMC run. We assume unbounded flat priors for the base cosmological parameters; the baryon density parameter $\Omega_{\rm b}$, the acoustic angular scale $100\theta_s$, the primordial comoving curvature power spectrum amplitude $\ln10^{10}A_s$ and the scalar spectral index $n_s$ while for the reionization optical depth $\tau$ and the initial dark matter density $\Omega_{\rm cdm}^{\rm ini}h^2\equiv \rho_{\rm cdm}^{\rm ini}a_\star^3 h^2/\rho_{\rm cr,0}$ \cite{2014JCAP...12..028A} we introduce a lower bound of $0.004$ and $10^{-9}$ respectively. In addition, for the decaying dark matter model parameters,  (decay rate $\Gamma$ and $\epsilon$), we assume flat priors in logarithmic space with  $-7<\log_{10}(\Gamma/{\rm Gyr}^{-1})<1$  and $-3<\log_{10}(\epsilon)<-0.3010$ -- recall that $\epsilon$ must always be less than $1/2$. Finally as an additional step to aid the numerical calculation and to attain convergence in a reasonable amount of time we use a $\Lambda$CDM covariance matrix as an input since our numerical tests did not show any effects of this choice to the results.

The main conclusion from \Fig{fig:fig5} is that {\it posteriors prefer a region nearly identical to that of} $\Lambda$CDM.  
The introduction of decaying dark matter has the effect of adding power at small multipoles (ISW effect) as well as increasing the amplitude of oscillations at high multipoles (CMB lensing -- see section II). What we see here, is that the predicted effects of two-body decaying dark matter on the CMB are heavily constrained by the observations of the CMB power spectrum. With $\Gamma$ and $\epsilon$ being limited to extremely small values, the decaying dark matter model becomes essentially degenerate with  $\Lambda$CDM -- for small $\Gamma$,  the majority of the parent dark matter particles do not decay, and  remain cold for the entire history of the Universe, while for small $\epsilon$, $m_2 \approx m_0$ effectively just relabelling the particles from $\psi$ to $\chi$ with very little radiation injected and resulting in no appreciable changes to the evolution. The effects discussed above are evident in \Fig{fig:fig5} with the dashed lines representing the best fit dark matter density $\Omega_m$  in  $\Lambda$CDM lying within the $68\%$ contours of the posterior distribution for the equivalent parameters in the decaying dark matter model. 

We find that $\epsilon$ is limited to small values, i.e., the mass difference between parent particle and massive daughter is small, since $\delta m = 1 - \sqrt{( 1 - 2 \epsilon)}$ (in units of parent particle mass), while $\Gamma$ is preferred to be $\Gamma \lesssim 1 \; {\mathrm{Gyr}^{-1}}$, i.e., lifetimes of order or greater the age of the Universe. 
For small values of $\epsilon $ the decay width remains unconstrained while the opposite is true for small values of $\Gamma$. We can approximate the $95\%$ confidence contour between these two parameters as $\epsilon \approx 0.002(\Gamma/{\mathrm{Gyr^{-1}}})^{-0.8}$. For $\epsilon\approx0.5$, our constraint on $\Gamma \lesssim 10^{-3} \; {\rm Gyr}^{-1}$ compares well with constraints placed on late time dark matter decays to radiation.~\cite{2014JCAP...12..028A} In the context of magnetic dipole transitions that lead to such decays (e.g., Super WIMPs or exited fermions that decay to a photon and a lighter fermion \cite{2003PhRvL..91a1302F, 2016PhRvD..94a5018C, 2018PhRvL.120v1102K}), the scale $\Lambda_\gamma$ of such process is related to the mass splitting between the parent and massive daughter particle $\delta m$ and the rate of decay through $\Lambda_\gamma \sim \sqrt{\delta m^3 / \Gamma}$. Since $\delta m$ is solely dependent on $\epsilon$ we can translate the degeneracy between $\epsilon$ and $\Gamma$ into a constraint on the scale as $\Lambda_\gamma \gtrsim 10^{13} \; \mathrm{GeV}$ for $\Gamma \lesssim 1 \; {\mathrm{Gyr}^{-1}}$.

Finally if we turn our attention to the two main quantities of interest, the Hubble parameter \H0, and the variance of the matter power spectrum fluctuations, \si8.  First, for \H0, the posterior median value (and 95\% intervals) in the case of two-body decays is $H_0 = 67.84 \pm 0.84 \,  \rm{km/s/Mpc}$. This shows clearly that the reshuffling of energy densities (from dark matter to radiation) in two-body decays {\it{cannot}} account for the speed-up of the expansion rate at late times as suggested in \cite{2019PhRvD..99l1302V}. Ref.~\cite{2019PhRvD..99l1302V} found a preferred region for these two parameters assuming a constant cosmology during CMB times and using late time observables. However, their work did not consider effects that occur to the CMB at late times like the ISW effect and lensing which we have shown here are very important. The preferred region in Ref.~\cite{2019PhRvD..99l1302V} is ruled out in our analysis, and remains such even with a combined analysis of {\it Planck} with an \H0 prior consistent with SH0es's measurements.~\cite{2019arXiv190307603R}

Similarly, for \si8, the posterior median (and 95\% intervals) is $\sigma_8 = 0.810 \pm 0.012$ ($S_8 = 0.822 \pm 0.021$), consistent with the value obtained in $\Lambda$CDM with  values of $67.82_{-0.82}^{+0.83} \, \rm{km/s/Mpc}$ and $0.810 \pm 0.012$ ($0.823 \pm 0.021$). Therefore, just as with \H0, we conclude that the introduction of the two-body decays cannot relieve the observed discrepancy between measurements of \si8. Here we have also presented $S_8= \sigma_8 \sqrt{\Omega_m/0.3}$ constraints for ease of comparison with other results.

To get a better understanding of the physical reasons behind the constraining power of the CMB power spectrum on two body decays, we can look closely into the physics of the decaying dark matter model. We have discussed in \Sec{sec:decaying_dm} how the effects of the decays scale with the parameter $\epsilon$. In order for decaying dark matter to be differentiable from CDM, $\epsilon$ must be large; otherwise, decays will not redistribute enough matter to radiation. In addition, the slow massive daughter particle effectively does not amount to any appreciable change in the evolution of the Universe.  The maximum ratio of energy transfer from matter to radiation is given by the mass difference between the parent and massive daughter particles, $ \delta m \sim 1-\sqrt{1-2\epsilon}$, which shows that for small $\epsilon$, $\delta m \approx \epsilon$. Even for the case of large decay width $\Gamma \gtrsim 1 \; \rm{Gyr}^{-1}$ for which most of the parent particles will have already decayed by today, the maximum allowed value for $\epsilon$ would be at best of order $10^{-2}$, well within the  Planck limits on percent level of variations off of a $\Lambda$CDM cosmology.

\section{Conclusion} \label{sec:conclusion}

In summary, we reviewed a decaying dark matter model of the form $\psi \rightarrow \gamma^\prime + \chi$, in the context of solving the \H0 and \si8 tensions. In the absence of CMB constraints, this model has been proposed as a solution to these cosmological problems because it leads to an increase in the expansion rate at late times and a decrease in the growth of structure \cite{2019PhRvD..99l1302V}. 
We find that this late decaying dark matter scenario is in dire straits due to the constraining power of the CMB anisotropies on low multiples (ISW effect) and on high multipoles (lensing), and we conclude that it cannot relieve neither the \H0 tension or \si8 tensions. These results are in agreement with the recent work of \cite{2020arXiv200407709H}.

A different decaying dark matter model was suggested in Ref.~\cite{2020arXiv200406114B} where instead of a decaying cold parent particle, it is a warm dark matter component that decays at around the time of matter - radiation equality. They showed that such a model is successful at reducing the tension between local and cosmological determinations of \H0. Other strong candidates remain like early dark energy \cite{2018arXiv181104083P,2020arXiv200413046C}, neutrino self interactions \cite{2019arXiv190200534K}, or any other proposal that shifts the epoch of recombination to earlier times. Time dependent DM properties also show promise in relieving the tension.~\cite{2020arXiv200409572I}

It seems however that no proposal so far has been uniquely successful in removing the two tensions. It is therefore imperative that more work is needed towards a solution together with new probes and data from future observations.

While our work was under review, \cite{2020arXiv200809615A} performed a similar analysis with a more rigorous treatment of the massive daughter perturbations. 

In both works (this work and \cite{2020arXiv200809615A}), there are minimal alterations to the preferences of \H0 and $S_8$, completely consistent with $\Lambda$CDM as stated above. 
However, \cite{2020arXiv200809615A} also performs an analysis that includes a prior on $S_8$. It is only when this additional constraint is included that there is an observed decrease of  $S_8$ ($S_8 = 0.795_{-0.015}^{+0.025}$). With the $S_8$ prior, we observe a decrease in the preferred value of $S_8$ as well, $S_8 = 0.810 \pm 0.009$, statistically consistent with the results of \cite{2020arXiv200809615A}. Therefore, it seems unlikely that late decaying dark matter solutions are viable explanations to the $H_0$ and $S_8$ tensions. 

\section{Acknowledgments}
We thank the anonymous referee for numerous comments that improved the work presented here. We acknowledge useful conversations with Manuel Buen-Abad, Jatan Buch, Isabelle Goldstein, Leah Jenks, John Leung,  Avi Loeb, Vivian Poulin, Adam Riess, and Michael Toomey. S.M.K  was partially supported by NSF-2014052. We gratefully acknowledge the support of Brown University. 

\bibliography{manuscript}

\end{document}